\documentclass[pra,nofootinbib,superscriptaddress,showpacs,showkeys]{revtex4}

\usepackage{graphicx,epsfig}
\usepackage{amsfonts,amsmath,amssymb,amsthm,amscd}

\begin{document}

\unitlength=1mm

\title{On $\mathbf{e^+e^-}$- pair production by a focused laser pulse in vacuum}

\author{S. S. Bulanov}
\email{bulanov@heron.itep.ru}
\affiliation{Institute of
Theoretical and Experimental Physics, 117218 Moscow, Russia}

\author{N.B. Narozhny}
\email{narozhny@theor.mephi.ru}
\author{V.D. Mur }
\affiliation{Moscow State Engineering Physics Institute, 115409
Moscow, Russia}

\author{V.S. Popov}
\affiliation{Institute of Theoretical and Experimental Physics,
117218 Moscow, Russia}

\begin{abstract}
The probability of electron-positron pair creation by a focused
laser pulse is calculated. For description of the focused laser
pulse we use a 3-dimensional model of the electromagnetic field
which is based on an exact solution of Maxwell equations. There
exists two types of focused waves: $e$- and $h$-polarized waves
with only either electric, or magnetic vector being transverse
respectively. It is shown that pair production is possible only in
$e$-polarized electromagnetic wave. The dependence of the pair
production probability on the intensity of the laser pulse is
obtained. It is argued that there exists a natural physical limit
for attainable focused laser pulse intensities. This limit is
posed by the pulse energy loss due to the effect of pair creation.
\end{abstract}



\maketitle

\section{Introduction}

The effect of electron-positron pair production in vacuum by a
strong electric field was first considered by Sauter \cite{Saut}
as early as in 1931, see also \cite{el-pos}. Schwinger was the
first who had derived the exact formula for the probability of
pair creation in a static electric field \cite{Schwinger} and
since then this process was often called the Schwinger effect.

The probability of pair creation acquires its optimum value when
the electric field strength is of the order of the "critical" for
QED value $E_S=m^2c^3/e\hbar=1.32\times 10^{16}$ V/cm. Clearly,
such field strength is unattainable for static fields
experimentally in near future. Therefore attention of many
researchers was focused on theoretical study of pair creation by
time-varying electric fields
\cite{B-I,3,4,NN,7,8,9,Ringwald,Popov}. Nowadays, a very strong
time-varying electromagnetic field can be practically realized
only with laser beams. First experiments dealing with nonlinear
QED effects caused by high energy electrons and photons
interacting with intense laser pulses have been carried out
recently. First, the observation of nonlinear Compton scattering
in the collision of $46.6$ GeV electron with a laser pulse of $
10^{18}$W/cm$^{2}$ intensity has been reported \cite {Bula}. Then,
the same group of researchers has observed electron-positron pair
creation in collision of laser photons, backscattered to GeV
energies by the $46.6$ GeV electron beam, with an intense laser
pulse \cite{Burke}. The results of the latter experiment were the
first laboratory evidence for inelastic light-by-light scattering
involving only real photons. However, the intensities of existing
now lasers are far lower than the value\footnote {We assume that
the laser beam is circularly polarized.}
$I_S=\displaystyle\frac{c}{4\pi}E_S^2\approx 4.6\times
10^{29}$W/cm$^{2}$ corresponding to the critical QED field $E_S$.
The estimations \cite{4,25,26} show that pair creation by a single
laser pulse in vacuum, or even in collision of two laser pulses,
could be hardly observed with lasers of intensity $I\ll I_S$.
Therefore the results of
Refs.~\cite{Saut,el-pos,Schwinger,B-I,3,4,NN,7,8} were commonly
believed to be of purely theoretical interest. Fortunately, the
latest achievements of laser technology promise very rapid growth
of peak laser intensities. Tajima and Mourou has suggested
recently \cite{Mourou} a path to reach an extremely high-intensity
level of $10^{26-28}$W/cm$^2$ already in the coming decade. Such
field intensities are very close to the critical value, $E_S$.
Hence the more detailed study of the Schwinger effect in
time-varying electromagnetic fields, in particular in the field of
a focused laser pulse, is becoming an urgent physical problem from
experimental point of view also.

As it was shown in Ref.~\cite{Schwinger}, a plane electromagnetic
wave of arbitrary intensity and spectral composition does not
create electron-positron pairs in vacuum because it has both field
invariants $ {\cal F}=({\bf E}^{2}-{\bf H}^{2})/2$ and ${\cal
G}=({\bf E\cdot }{\bf H})$ equal to zero. Therefore we consider
the effect in the field of a focused circularly polarized laser
pulse which is described by a realistic 3-dimensional model
developed in Ref.~\cite{NB}. Unlike the case of spatially
homogeneous time-varying electric field used in
Refs.~\cite{B-I,3,4,NN,7,8,9,Ringwald,Popov}, the utilized model
is based on an exact solution of Maxwell equations. The model has
been already successfully used in Ref~\cite{NF} for quantitative
explanation of anisotropy of electrons accelerated by a
high-intensity laser pulse which was observed in experiment of
Malka {\it et al} \cite{MLM}.

The method we use in the present paper for calculation of the
number of pairs created by the laser pulse is based on the fact
that the characteristic length of the process is determined by the
Compton length $l_c=\hbar/mc$ which is much less than the
wavelength $\lambda$ of the laser field, $\,l_c\ll\lambda\,$.
Therefore, at arbitrary point of the pulse we can calculate the
number of created particles per unit volume and unit time
according to the Schwinger formula for the static homogeneous
field and then obtain the total number of created particles as the
integral over the volume $V$ and duration $\tau$ of the pulse
\begin{equation}\label{NT}
N=\frac{e^2E_S^2}{4\pi^2 \hbar^2 c} \int\limits_V
dV\int\limits_0^\tau dt\; \epsilon\eta\coth\frac{\pi
\eta}{\epsilon} \exp\left(-\frac{\pi}{\epsilon}\right).
\end{equation}
Here $\epsilon={\mathcal{E}}/E_S,\, \eta={\mathcal{H}}/E_S$ are
the reduced fields, and $\mathcal{E}$ and $\mathcal{H}$ are the
field invariants which have the meaning of electric and magnetic
fields in the reference frame where they are parallel. The
invariants $\mathcal{E}$ and $\mathcal{H}$ can be expressed in
terms of invariants $\mathcal{F}$ and $\mathcal{G}$, see, e.g.,
Ref.~\cite{Schwinger},
\begin{equation}\label{EH}
\begin{array}{c}
\displaystyle {\mathcal{E}}=\sqrt{\left({\cal F}^2+{\cal
G}^2\right)^{1/2}+{\cal F}}, ~~~ \displaystyle
{\mathcal{H}}=\sqrt{\left({\cal F}^2+{\cal G}^2\right)^{1/2}-{\cal
F}}.
\end{array}
\end{equation}

The paper is organized as follows. In Sec.~II we consider the
model of the focused laser pulse field and calculate the field
invariants. The qualitative discussion of the pair production
process is carried out in Sec.~III. We present the results of
numerical calculations of the pair production probability in
Sec.~IV. The summary of the results and conclusions are presented
in Sec.~V.

\section{Model of the field.}

It is well known that the electromagnetic field of a focused light
beam is not transverse. Therefore, strictly speaking, one cannot
ascribe some definite type of polarization to it. However, we can
always represent the field of a focused beam as a superposition of
fields with transverse either electric, or magnetic vector only,
see, e.g., Ref.~\cite{BW}. For each of these fields we can define
the type of polarization with respect to that vector which is
transverse. We will call such fields $e$- or $h$-polarized fields
respectively.

It can be verified straightforwardly that there exists the
following exact solution of Maxwell equations which describes a
wave propagating along the $z$ axis  \cite{NB}
\begin{equation}\label{E_e}
{\bf{E}}^e=iE_0  e^{-i\varphi}\left\{ F_1({\bf e}_x\pm i{\bf
e}_y)-F_2 e^{\pm 2 i\phi}({\bf e}_x\mp i{\bf e}_y)\right\},
\end{equation}
\begin{equation}\label{H_e}{\bf{H}}^e=\pm E_0 e^{-i \varphi}
\left\{\left(1-i\Delta^2\frac{\partial}{\partial\chi}\right)
\left[F_1({\bf e}_x\pm i{\bf e}_y)+F_2 e^{\pm 2 i\phi}({\bf
e}_x\mp i{\bf e}_y)\right]+2i\Delta e^{\pm i\phi}\frac{\partial
F_1}{\partial \xi}{\bf e}_z \right\}.
\end{equation}
Here $\omega$ is the wave frequency, $x$, $y$, and $z$ are spatial
coordinates, and
\begin{equation}
\begin{array}{c}
\varphi=\omega(t-z/c),~~~\xi=\rho/R,~~~\chi=z/L,  \\ \\
\rho=\sqrt{x^2+y^2},~~~\cos\phi=x/\rho,~~~\sin\phi=y/\rho, \\ \\
\Delta\equiv c/\omega R=\lambda/2\pi R,~~~L\equiv R/\Delta\,.
\end{array}
\end{equation}
The fields (\ref{E_e}) and (\ref{H_e}) are solutions of Maxwell
equations if function $F_1(\xi,\chi;\Delta)$ obeys the equation
\cite{NB}
\begin{equation}\label{F_1}
2i\frac{\partial F_1}{\partial\chi}+\Delta^2\frac{\partial^2
F_1}{\partial\chi^2}+\frac{1}{\xi}\frac{\partial}{\partial\xi}
\left(\xi\frac{\partial F_1}{\partial\xi}\right)=0\,,
\end{equation}
and function $F_2(\xi,\chi;\Delta)$ is defined by the relation
\begin{equation}\label{F_2}
F_2(\xi)=F_1(\xi)-\frac{2}{\xi^2}\int\limits_0^{\xi} \xi^\prime
F_1(\xi^\prime)d\xi^\prime.
\end{equation}
The fields (\ref{E_e}), (\ref{H_e}) describe a focused beam if
functions $F_1,F_2$ are chosen so that they tend to zero
sufficiently fast when $\xi,|\chi|\rightarrow\infty$ and satisfy
the following conditions
\begin{equation}\label{b_c1}
\lim_{\Delta\rightarrow 0}F_1(0,0;\Delta)=1,\quad
\lim_{\Delta\rightarrow 0}F_2(0,0;\Delta)=0\,.
\end{equation}
In this case the parameter $R$ can be interpreted as the radius of
the focal spot, $L$ as the diffraction length \cite{NB}, and
$\Delta$ as the focusing parameter. If $\Delta=0$ the solution
(\ref{E_e}), (\ref{H_e}) describes a plane wave.

Even if the laser beam is focused so that $R\sim\lambda$ (this
corresponds to the diffraction limit), $\Delta\sim10^{-1}$.
Therefore in our calculations we will always assume that
$\Delta\ll 1$. Under this assumption the field (\ref{E_e}) in the
spatial area $$\xi \ll1,\quad |\chi|\ll 1\,,$$ in virtue of
Eqs.~(\ref{b_c1}), is very close to electric field of a circularly
polarized plane wave. In this sense we will identify the field
(\ref{E_e}), (\ref{H_e}) with circularly $e$-polarized focused
light beam. The electric and magnetic fields in the circularly
$h$-polarized beam are given by the following expressions
\cite{NB}
\begin{equation}
{\bf{E}}^h=\pm i {\bf{H}}^e,~~~{\bf{H}}^h=\mp i {\bf{E}}^e.
\label{H-wave}
\end{equation}
For discussion of other types of polarizations of focused light
beams see Ref.~\cite{NB}.

The exploited model admits different field configurations, which
are determined by two functions $F_1, F_2$. One of the possible
variants for solutions of Eqs.~(\ref{F_1}), (\ref{F_2}) satisfying
constrains (\ref{b_c1}) can be expressed at $\Delta\ll 1$ as
\begin{equation}\label{G_b}
\begin{array}{c}
\displaystyle
F_1=(1+2i\chi)^{-2}\left(1-\frac{\xi^2}{1+2i\chi}\right)
\exp\left(-\frac{\xi^2}{1+2i\chi}\right),\\ \\
\displaystyle
F_2=-\xi^2(1+2i\chi)^{-3}\exp\left(-\frac{\xi^2}{1+2i\chi}\right)\,,
\end{array}
\end{equation}
see Ref.~\cite{NB}. The waves of such type are commonly called
Gaussian beams. We will work with expressions (\ref{G_b}) for
functions $F_1, F_2$ throughout the paper.

To describe a laser pulse with finite duration $\tau$ one should
introduce \cite{NB} a temporal amplitude envelope
$g(\varphi/\omega\tau)$ making the following substitutions in
Eqs.~(\ref{E_e}), (\ref{H_e})
\begin{equation}\label{subst}
\exp(-i\varphi) \rightarrow if'(\varphi),\quad
\Delta\exp(-i\varphi)\rightarrow \Delta f(\varphi),
\end{equation}
where $$f(\varphi)=g(\varphi/\omega\tau)\exp(-i\varphi)\,.$$ It is
assumed that the function~$g(\varphi/\omega\tau)$ is equal to
unity at the point~$\varphi=0$ and decreases exponentially at the
periphery of the pulse for $|\varphi|\gg\omega\tau$. In this case
the electric and magnetic fields of the model constitute an
approximate solution of Maxwell equations with the second-order
accuracy with respect to small parameters $\Delta$ and
$\Delta^\prime=1/\omega\tau$
\begin{equation}
\label{1} \Delta^\prime\lesssim\Delta \ll 1\,.
\end{equation}

Using Eqs.~(\ref{E_e}), (\ref{H_e}) we can calculate the field
invariants ${\cal F},\,{\cal G}$. For the sake of compactness we
will give here explicit expressions for them only for the case of
$e$-polarized pulses. In the lowest order with respect to
parameter $\Delta$ they read
\begin{equation}\label{INVS}
\begin{array}{c}
\displaystyle {\cal F}^e=\frac{1}{2}\left\{({\rm
Re}{\bf{E}}^e)^2-({\rm Re}{\bf{H}}^e)^2 \right\}=\\ \\
\displaystyle =2E_0^2g^2(\varphi/\omega\tau) \Delta^2\left\{{\rm
Im}\left[F_1\frac{\partial
F_1^\ast}{\partial\chi}+F_2\frac{\partial
F_2^\ast}{\partial\chi}\right]-\left|\frac{\partial F_1
}{\partial\xi}\right|^2 +{\rm
Re}\left[e^{-2i(\varphi\mp\phi)}\left(\left(\frac{\partial
F_1}{\partial\xi}\right)^2+i\frac{\partial}{\partial
\chi}\left(F_1 F_2\right)\right)\right]\right\}, \\
\\ \displaystyle {\cal G}^e={\rm Re}{\bf{
E}}^e{\rm Re}{\bf{H}}^e=\pm 2E_0^2g^2(\varphi/\omega\tau)
\Delta^2\left\{{\rm Re}\left(F_2\frac{\partial F_2^\ast}{\partial
\chi}-F_1\frac{\partial F_1^\ast}{\partial \chi}\right)-{\rm
Re}\left[\left(F_2\frac{\partial F_1^\ast}{\partial
\chi}-F_1\frac{\partial F_2^\ast}{\partial
\chi}\right)e^{-2i\varphi\pm 2 i\phi}\right] \right\}.
\end{array}
\end{equation}
As it should be, at $\Delta=0$, i.e. in the case of plane wave
field, both invariants are equal to zero. The invariants
${\mathcal{E}},\,{\mathcal{H}}$ can be calculated by substitution
of Eqs.~(\ref{INVS}) into Eqs.(\ref{EH}).

\section{Qualitative discussion.}

Now, using the model described in the preceding section, we will
give a qualitative estimate of the number of pairs created by a
focused laser pulse in vacuum. Taking into account that pairs are
created mainly near the focus, we assume that the spatial volume
of the pulse is of the order of $\pi R^2c\tau$ and the number of
created pairs is given by the following expression
\begin{equation}\label{N_est}
N\approx\frac{e^2E_S^2}{4\pi^2 \hbar^2 c} \pi R^2c\tau^2
\overline{\epsilon}\,\overline{\eta}\coth\frac{\pi
\overline{\eta}}{\overline{\epsilon}}
\exp\left(-\frac{\pi}{\overline{\epsilon}}\right),
\end{equation}
instead of Eq.~(\ref{NT}). Here $\overline{\epsilon}$ and
$\overline{\eta}$ are the averaged over time values of
dimensionless field invariants $\epsilon$ and $\eta$ in the focus.

Using Eqs.~(\ref{INVS}), (\ref{G_b}) we easily find that in the
focus $\xi=0,\,\chi=0$ invariants $\mathcal{F}$ and $\mathcal{G}$
in the lowest approximation with respect to $\Delta$ are equal to
\begin{equation}\label{inv_f}
\begin{array}{c}
\displaystyle{\cal F}^e(0,0)=8\Delta^2E_0^2
g^2\left(\frac{t}{\tau}\right),~~~{\cal G}^e(0,0)=0, \\ \\
\displaystyle{\cal F}^h(0,0)=-8\Delta^2E_0^2
g^2\left(\frac{t}{\tau}\right),~~~{\cal G}^h(0,0)=0.
\end{array}
\end{equation}
It immediately follows from Eqs.~(\ref{EH}) that at the focus of
the $e$-polarized pulse
\begin{equation}\label{eps_e} \epsilon=(2{\cal
F}^e/E_S^2)^{1/2},\quad \eta=0\,,
\end{equation}
while for the $h$-polarized pulse we have
\begin{equation}\label{eps_h}\epsilon=0,\quad \eta =(2{\cal
F}^e/E_S^2)^{1/2}\,.
\end{equation}
So we conclude that pairs can be created only by a $e$-polarized
but not $h$-polarized pulse. This statement is not exact however.
We will see in the next section that $h$-polarized pulse also
create pairs though their number is several orders of magnitude
less then in the case of $e$-polarized pulse.

Now we can estimate the number $N$ of pairs created by a
$e$-polarized focused pulse. However, it is convenient to express
$N$ in terms of laser intensities instead of
$\overline{\epsilon}$. To do this, we begin from the Pointing
vector averaged over fast oscillations of electromagnetic field of
the pulse
\begin{equation}
<{\bf S}>=\frac{c}{4\pi}E_0^2
g^2\left(\frac{\varphi}{\omega\tau}\right)
\left\{|F_1|^2+|F_2|^2\right\}{\bf e}_z\,.
\end{equation}
Using explicit expressions for functions $F_1,\,F_2$ (\ref{G_b}),
we obtain for the energy flow through the focal plane
\begin{equation}
\Phi=\int<S_z>\big|_{z=0}~dx~dy=\frac{c}{4\pi}E_0^2
g^2\left(\frac{\varphi}{\omega\tau}\right) \frac{\pi R^2}{2}\,.
\end{equation}
The total energy carried by the pulse is
\begin{equation}\label{W1}
W=\int\limits_{-\infty}^{\infty}dt\,\Phi=G\frac{c}{4\pi}E_0^2\pi
R^2\tau \,,
\end{equation}
where $$G=\frac{1}{2}\int\limits_{-\infty}^{\infty}du\,
g^2(u)\,,$$ and $\tau$ is the pulse duration. Further on we will
use the gaussian temporal amplitude envelope $\displaystyle
g\big|_{z=0}=e^{-4t^2/\tau^2}$, for which $\displaystyle
G=\frac{1}{4}\sqrt{\frac{\pi}{2}}\approx 0.31\,.$ By definition,
the intensity $I$ is
\begin{equation}\label{W2}
I=W/\pi
R^2\tau=G\frac{c}{4\pi}E_0^2
\,.
\end{equation}
The factor $G$ in the right-hand side of the latter equality has
arisen due to the fact that the mean value of the electric field
in the pulse is less than its peak value $E_0$. For the averaged
over time invariant $\overline{\epsilon}$ we get using
Eqs.~(\ref{eps_e}) and (\ref{inv_f})
\begin{equation}\label{eps_a}
\overline{\epsilon}=\frac{1}{\tau}\int\limits_{-\infty}^{\infty}
\epsilon dt=2\sqrt{\pi}\Delta \frac{E_0}{E_S}\,.
\end{equation}
Finally we obtain from Eqs.~(\ref{W2}), (\ref{eps_a})
\begin{equation}\label{eps_f}
\frac{I}{I_S}=\frac{1}{\sqrt{2\pi}}\frac{{\overline{\epsilon}}\,^2}{16\Delta^2}\,.
\end{equation}

Let us estimate the number of pairs created by the $e$-polarized
pulse with critical peak electric field $E_0=E_S$ for
$\Delta=0.1$, $\lambda=1~\mu m$ and $\tau=10$ fs. In this case
$\overline{\epsilon}\approx 0.35, \; I\approx 0.31\,I_S\approx
1.5\times 10^{29}$W/cm$^2$\, and according to Eq.~(\ref{N_est})
$N_e\sim 5\cdot 10^{20}$. If intensity of the same pulse is equal
to critical value $I=I_S$ we have $\overline{\epsilon}\approx
0.63$, and $N_e\sim 10^{23}$. We see that the number of created
pairs grows very fast with intensity. Indeed, only three times
increase of intensity yields three orders of magnitude in the
number of created pairs.

Let us now compare the rest energy of created pairs with the total
energy carried by the laser pulse which according to
Eq.~(\ref{W1}) for accepted values of parameters
$\Delta,\,\lambda$ and $\tau$ is given by the following formula
$$W\approx 5\cdot 10^{21}\frac{I}{I_S}mc^2.$$ We see that the energy of
the pulse with intensity $I\approx 0.31I_S$ is of the order $W\sim
10^{21}mc^2$ and hence is of the same order as the rest energy of
created pairs $W_p\sim 10^{21}mc^2$. Thus, we conclude that the
exploited method becomes inconsistent and one should take into
account back reaction of the pair creation effect on the process
of laser pulse focusing at such intensity. In other words, one
cannot consider the electromagnetic field of the pulse at near
critical intensities as a given external field and should take
into account depletion of the pulse due to pair production.
Moreover, it is clear that the critical intensity $I_S$ can hardly
be achieved for at least the $e$-polarized focused laser pulse in
optical range. We will discuss the situation with $h$-polarized
pulses in Sec.~V.

To conclude this section, we should emphasize that it follows from
Eq.~(\ref{N_est}) that the number of created pairs $N_e\approx 30$
at $I=5\cdot 10^{27}$W/cm$^2$ for the same values of parameters
$\Delta,\,\lambda$ and $\tau$ in $e$-polarized pulse. Hence we can
say that the effect of pair creation becomes observable just at
intensity of such order of magnitude. The peak value of electrical
field for the pulse of such intensity is of the order of $E_0\sim
0.18E_S$, $\overline{\epsilon}\sim 6\cdot 10^{-2}$ and the rest
energy of created pairs $W_p\sim 4mc^2\ll W\approx 5\cdot
10^{19}mc^2 $. The last inequality approves the application of our
method to the considered problem at $I=5\cdot 10^{27}$W/cm$^2$.

\section{Numerical calculations.}

In this section we present the results of numerical calculations
of the number of created pairs (\ref{NT}) by the fields
(\ref{E_e}), (\ref{H_e}).

Figs.~1, 2 demonstrate the dependencies of invariants
$\mathcal{E}$ and $\mathcal{H}$ for a $e$-polarized wave on
spatial coordinates $x$ and $y$ for the time moments $t=0,\,
t=\pi/2\omega$ respectively.
\begin{figure}[ht]
\begin{tabular}{ccc}
\epsfxsize8cm\epsffile{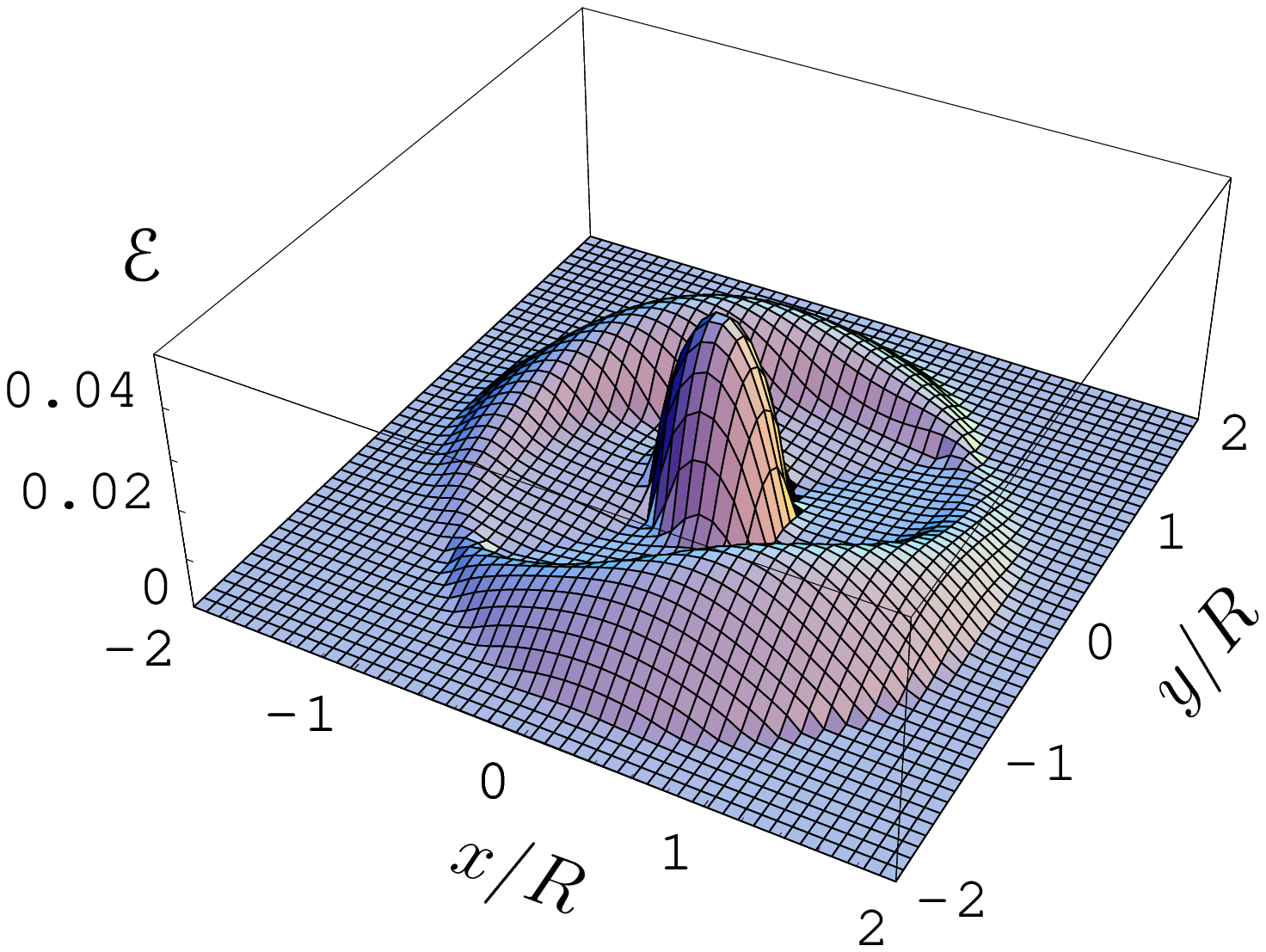} &
\epsfxsize8cm\epsffile{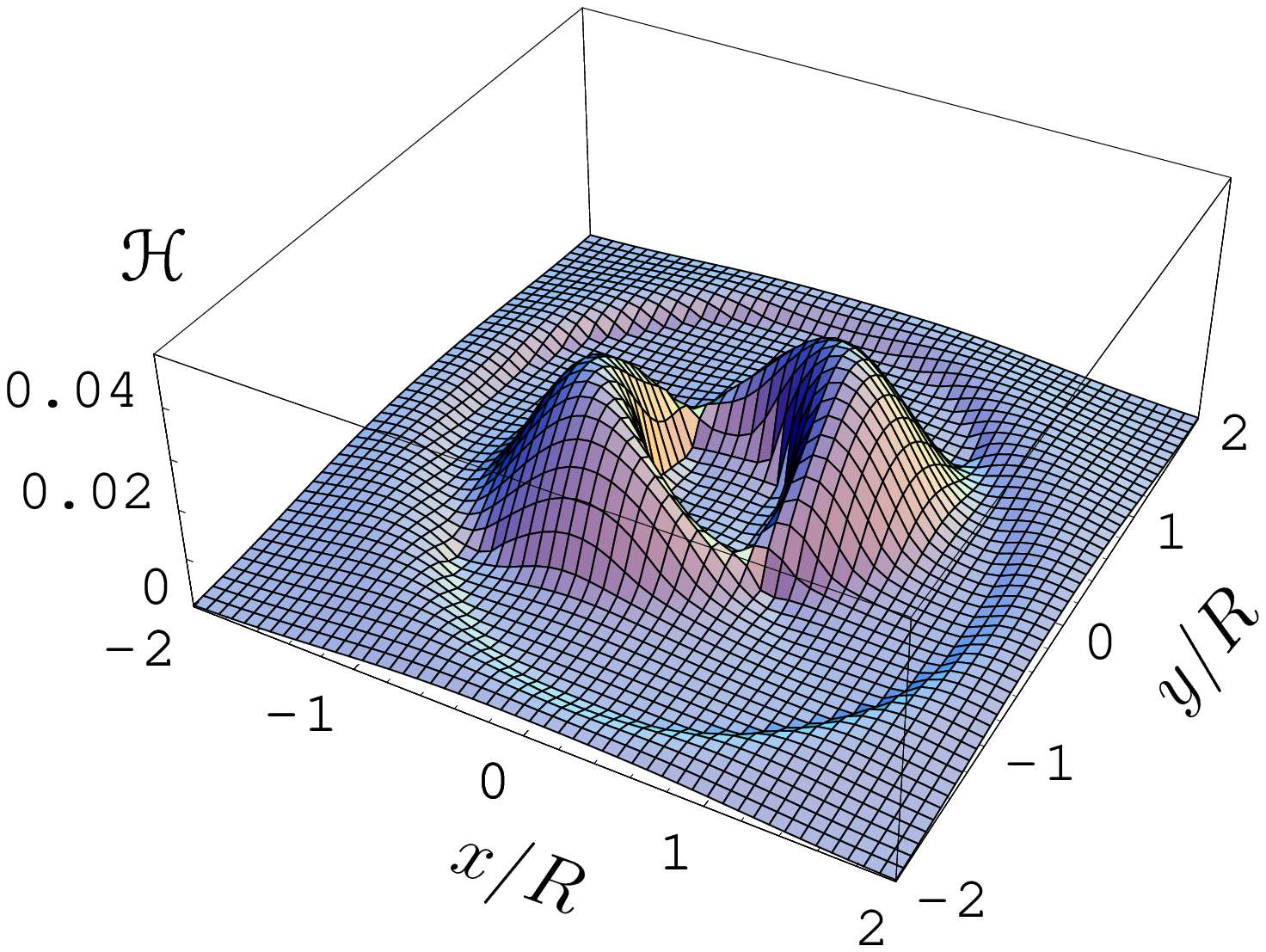} &
\\ a) & b) &
\end{tabular}
\caption{The dependencies of $\mathcal{E}$ (a) and $\mathcal{H}$
(b) on spatial coordinates $x$ and $y$ for the time moment $t=0$.
$\mathcal{E}$ and $\mathcal{H}$ are measured in units of $E_S$,
and the other parameters are chosen $E_0=0.1, ~~z=0,
~~\Delta=0.1.$}
\end{figure}
\begin{figure}[ht]
\begin{tabular}{ccc}
\epsfxsize8cm\epsffile{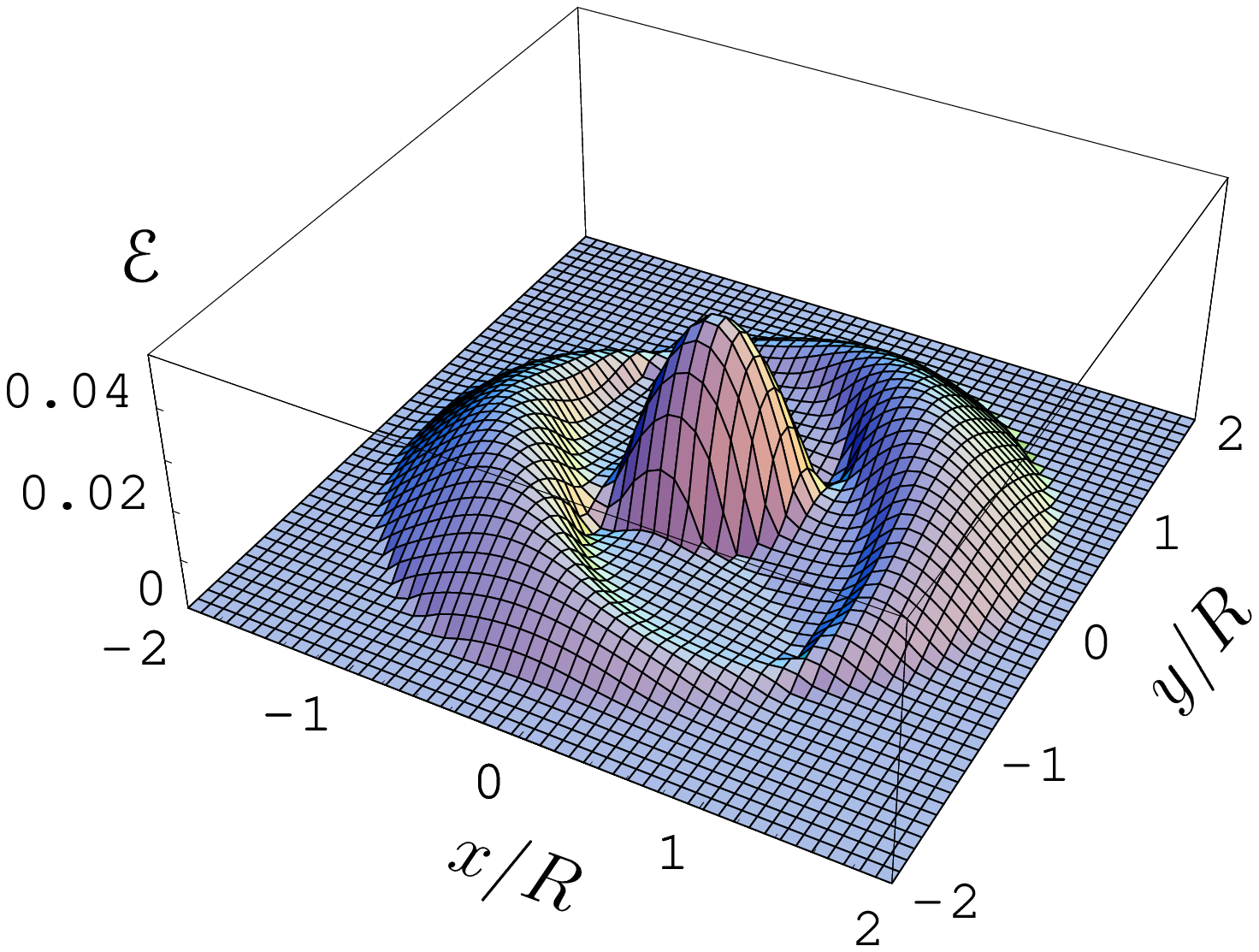} &
\epsfxsize8cm\epsffile{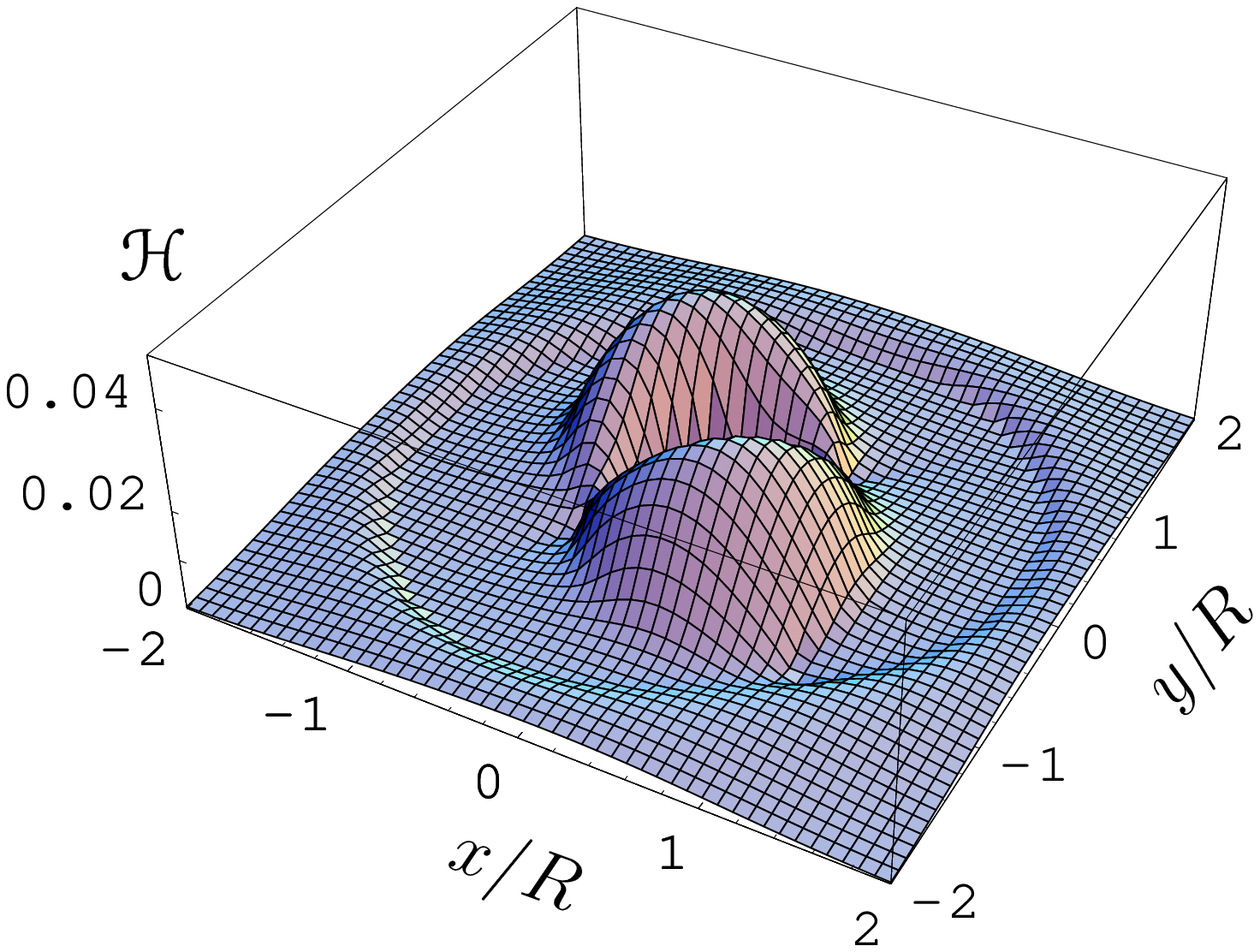} &
\\ a) & b) &
\end{tabular}
\caption{$\mathcal{E}$ (a) and $\mathcal{H}$ (b) as functions of
spatial coordinates $x$ and $y$ for the time moment
$t=\pi/2\omega$. The parameters are the same as in Fig.~1.}
\end{figure}
It can be seen from these figures that the electric field
$\mathcal{E}$ reaches its maximum in the focus ($x=0$, $y=0$),
while the maximum of the field $\mathcal{H}$ moves with time along
the circle of radius $\xi\approx 0.6$, and its value is less than
the value of $\mathcal{E}$ maximum.

In Fig. 3 we present the dependencies of invariants $\mathcal{E}$
and $\mathcal{H}$ on spatial coordinate $\chi$. One can see that
$\mathcal{E}$ has a maximum in the focus, while $\mathcal{H}$ is
very close to zero there. The forms of $\mathcal{E}$ and
$\mathcal{H}$ dependencies on spatial coordinates as well as on
time (which can be deduced from comparing Fig.~1 and Fig.~2)
justify the method of estimation used in the preceding section,
which is based on the assumption that pairs are produced mainly
near the focus where the values of $\mathcal{E}$ and $\mathcal{H}$
are independent of time apart the factor $g(\varphi/\omega\tau)$.
\begin{figure}[b]
\begin{tabular}{ccc}
\epsfxsize8cm\epsffile{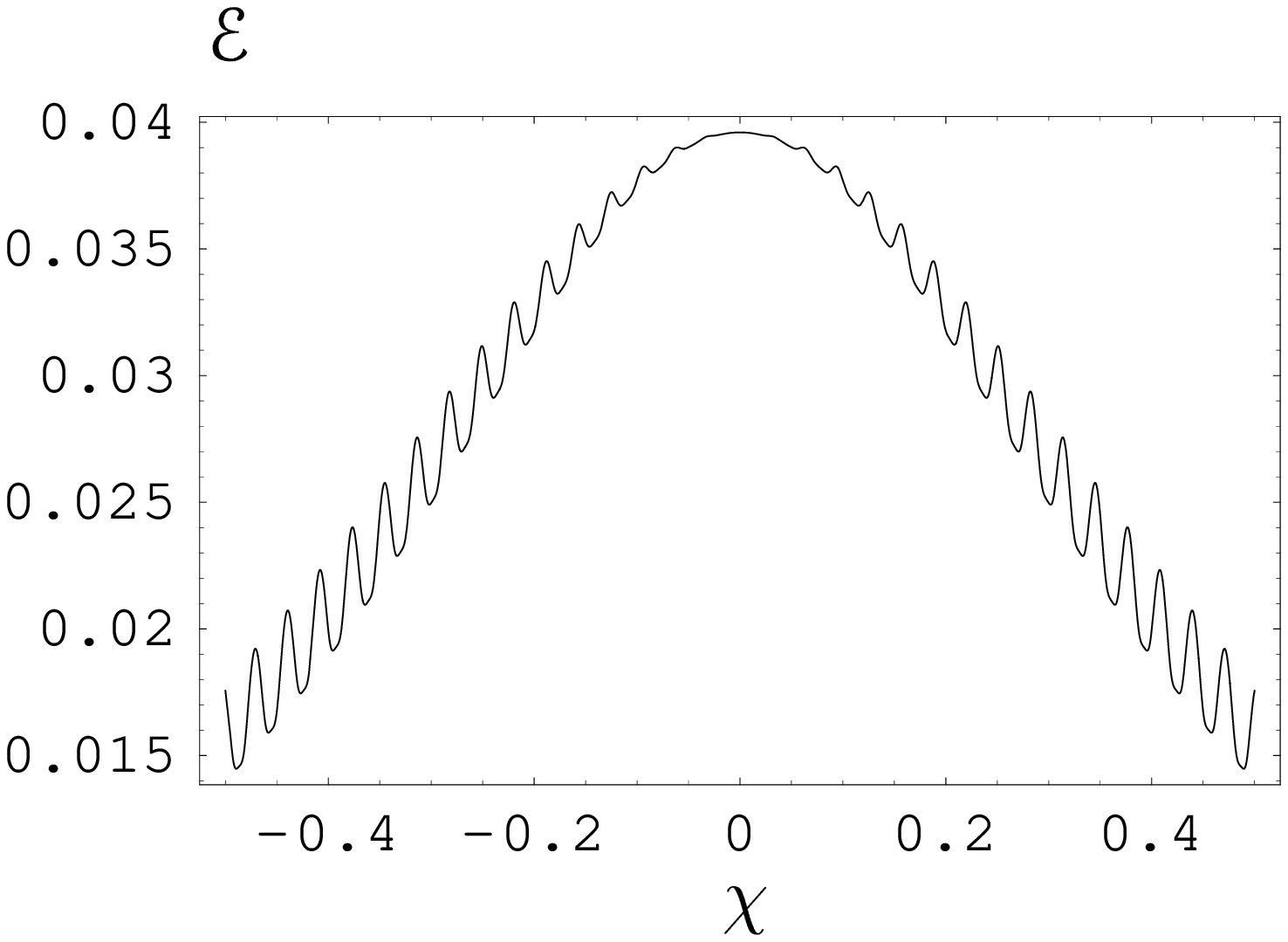} &
\epsfxsize8cm\epsffile{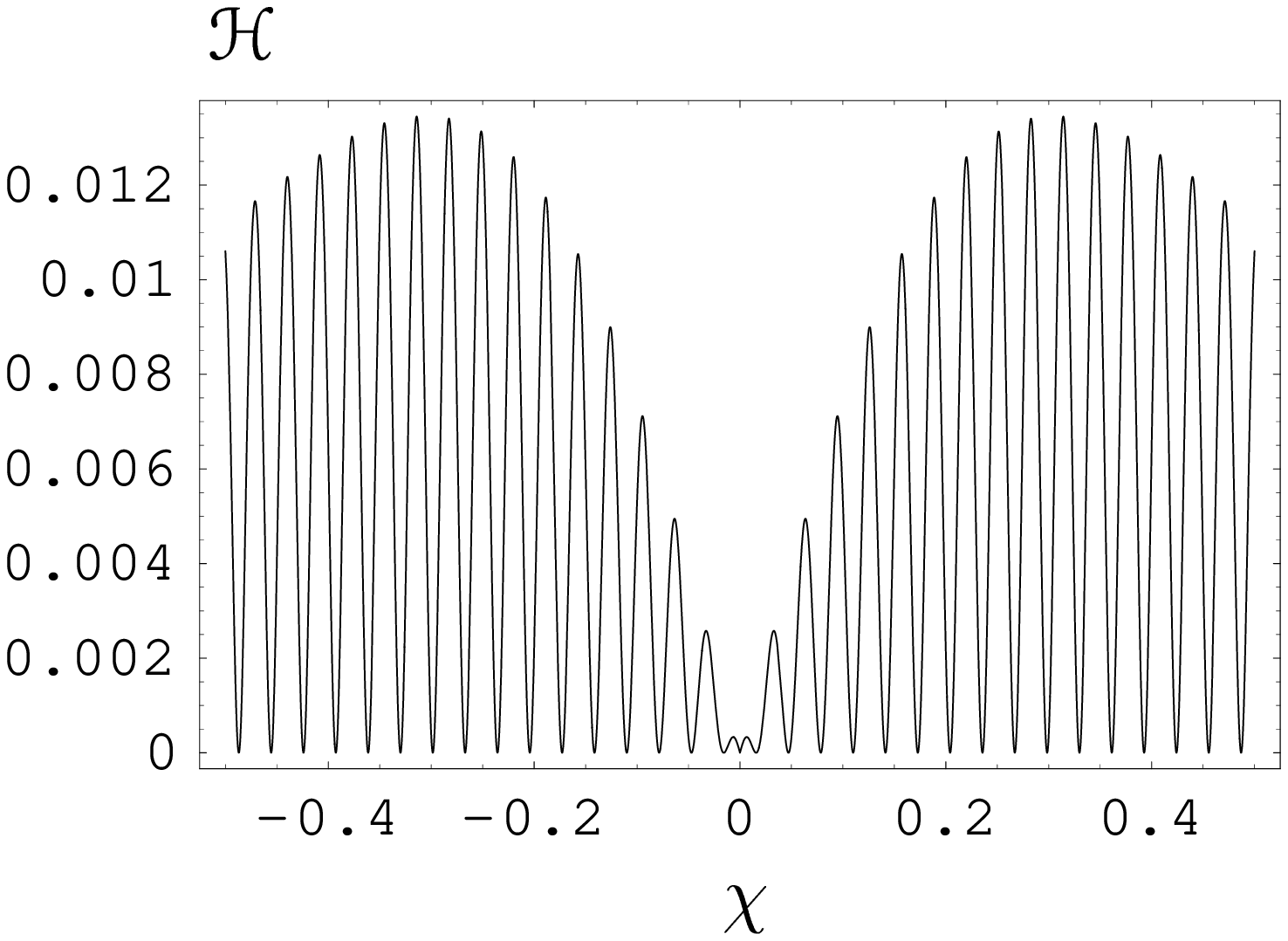} &
\\ a) & b) &
\end{tabular}
\caption{The dependencies of $\mathcal{E}$ (a) and $\mathcal{H}$
(b) on spatial coordinate $\chi=z/L\,$ for the time moment $t=0$.
$\mathcal{E}$ and $\mathcal{H}$ are measured in units of $E_S$,
and the other parameters are chosen $E_0=0.1, ~~x=0, ~~y=0,
~~\Delta=0.1.$}
\end{figure}

The coordinate dependencies of the invariants $\mathcal{E}$ and
$\mathcal{H}$ in the $h$-polarized wave are also given by Figs.~1-
2 if one makes the interchange
${\mathcal{E}}\leftrightarrows{\mathcal{H}}$. These results
explain why we have obtained the zero value for the number of
created pairs by the $h$-polarized wave. Indeed, our method of
estimation used in the preceding section was based on the value of
the invariants in the focus and, as it is clear from the above,
$\mathcal{E}$ is equal there to zero for the $h$-polarized wave.
In fact, $\mathcal{E}$ is not equal to zero at the periphery of
the focal plane. However its values are suppressed by the exponent
in functions $F_1, \,F_2$, see Eq.~(\ref{G_b}). Hence the number
of pairs produced by an $h$-polarized wave is several orders of
magnitude less than by the $e$-polarized wave of the same
intensity and parameter $\Delta$, see below.

\begin{figure}[ht]
\begin{tabular}{ccc}
\epsfxsize8cm\epsffile{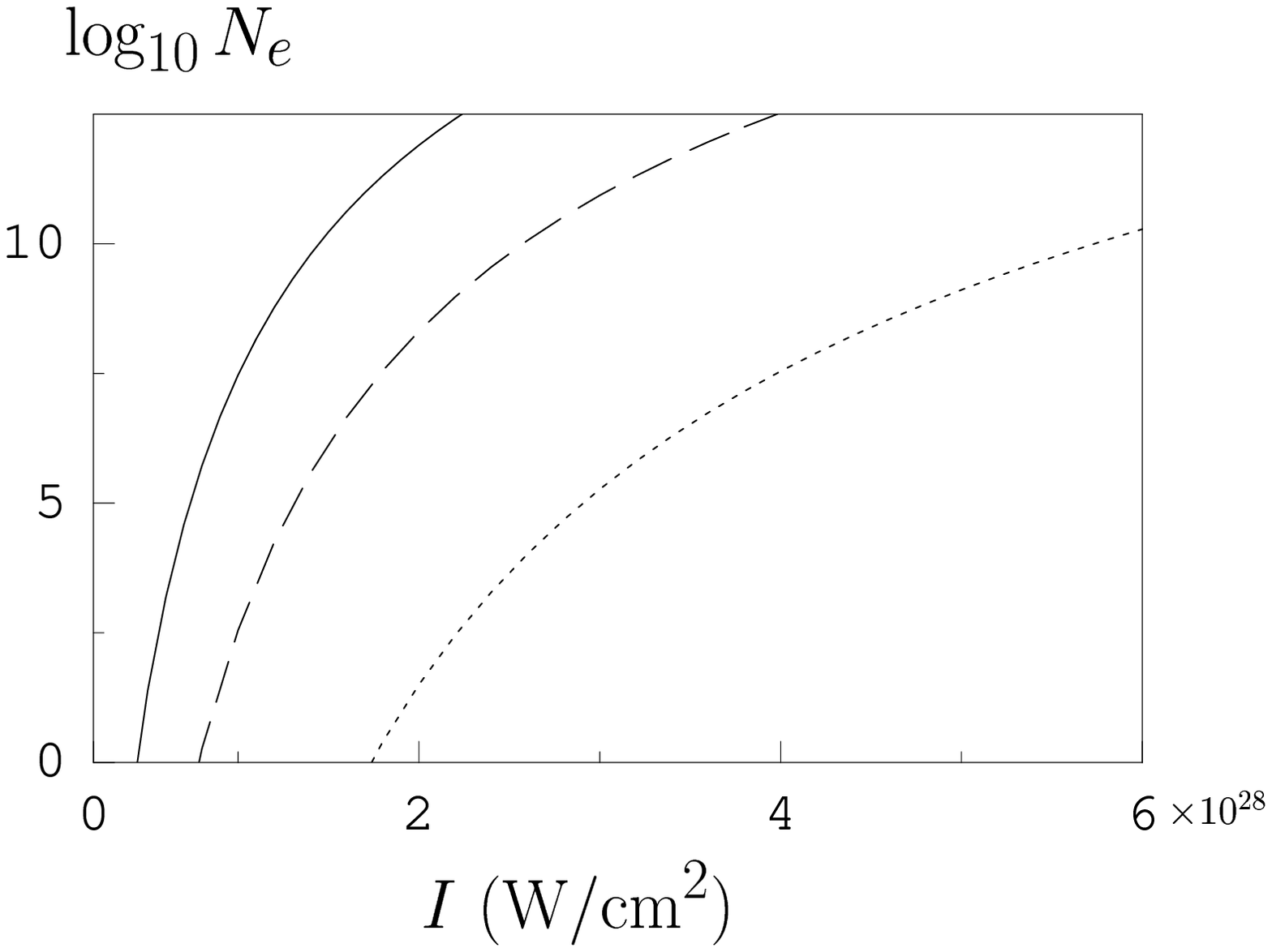} &
\epsfxsize8cm\epsffile{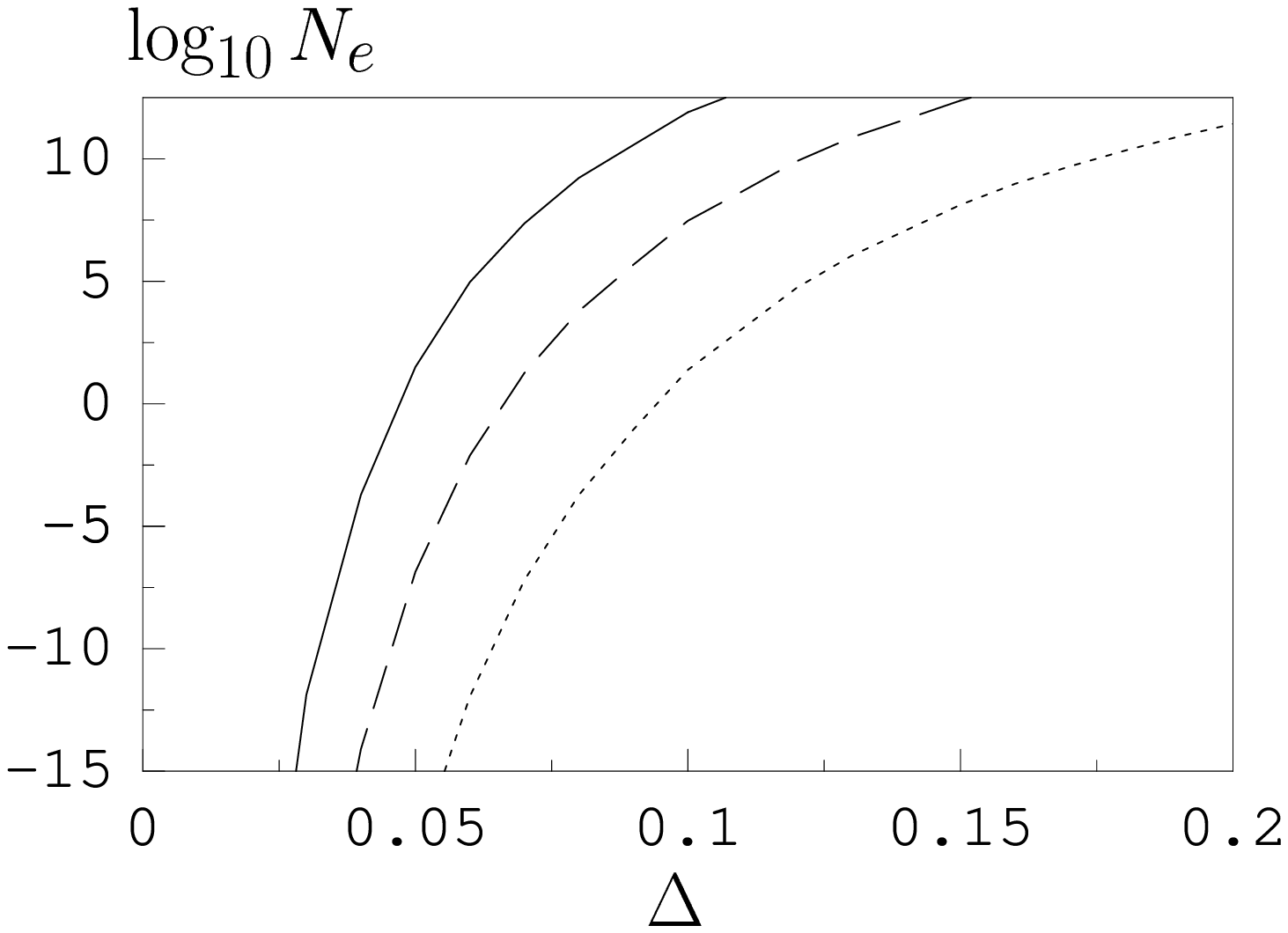} &
\\ a) & b) &
\end{tabular}
\caption{a) The dependence of the number of created pairs on laser
pulse intensity for different values of $\Delta$
($\Delta=0.1,~0.075,~0.05$ from top to bottom) and for $\lambda=1
\mu m$, $\tau=10^{-14}$ s. b) The dependence of the number of
created pairs on $\Delta$ for different values of laser pulse
intensity ($I=0.5\times 10^{28}$ W/cm$^2$, $1\times 10^{28}$
W/cm$^2$, $2\times 10^{28}$ W/cm$^2$ from bottom to top)}
\end{figure}
In Fig.~4a we present the dependence of the number $N_e$ of pairs
created by a $e$-polarized laser pulse on intensity for different
values of $\Delta$ and for $\lambda=1 \mu m$, $\tau =10^{-14}$ s.
We see that the number of created pairs grows very rapidly when
intensity increases from $10^{27}$ W/cm$^2$ to $10^{28}$ W/cm$^2$.
In agreement with out estimation the number of created pairs
reaches the value of the order $1$ at intensity close to $5\cdot
10^{27}$ W/cm$^2$. This means that the effect of pair creation can
be experimentally observed for the laser pulse intensity two
orders of magnitude less the critical value $I_S$. This is true
for at least a $e$-polarized pulse and can be explained by the
large value of the preexponential factor in (\ref{N_est}), which
can be represented as the ratio of the laser pulse 4-volume to the
Compton 4-volume. This factor compensates small Schwinger exponent
in the expression for the number of created pairs.

Fig.~4b illustrates the dependence of the number of created pairs
on the parameter $\Delta$. One can see that for all values of
intensity the number of pairs very rapidly approaches zero with
decrease of $\Delta$. It is due to the fact that the less is
$\Delta$, the better the focused laser pulse can be approximated
by a monochromatic plane wave which cannot create pairs at all.
This can be explicitly seen from the expressions (\ref{INVS}) for
the field invariants.

\begin{figure}[ht]
\epsfxsize8cm\epsffile{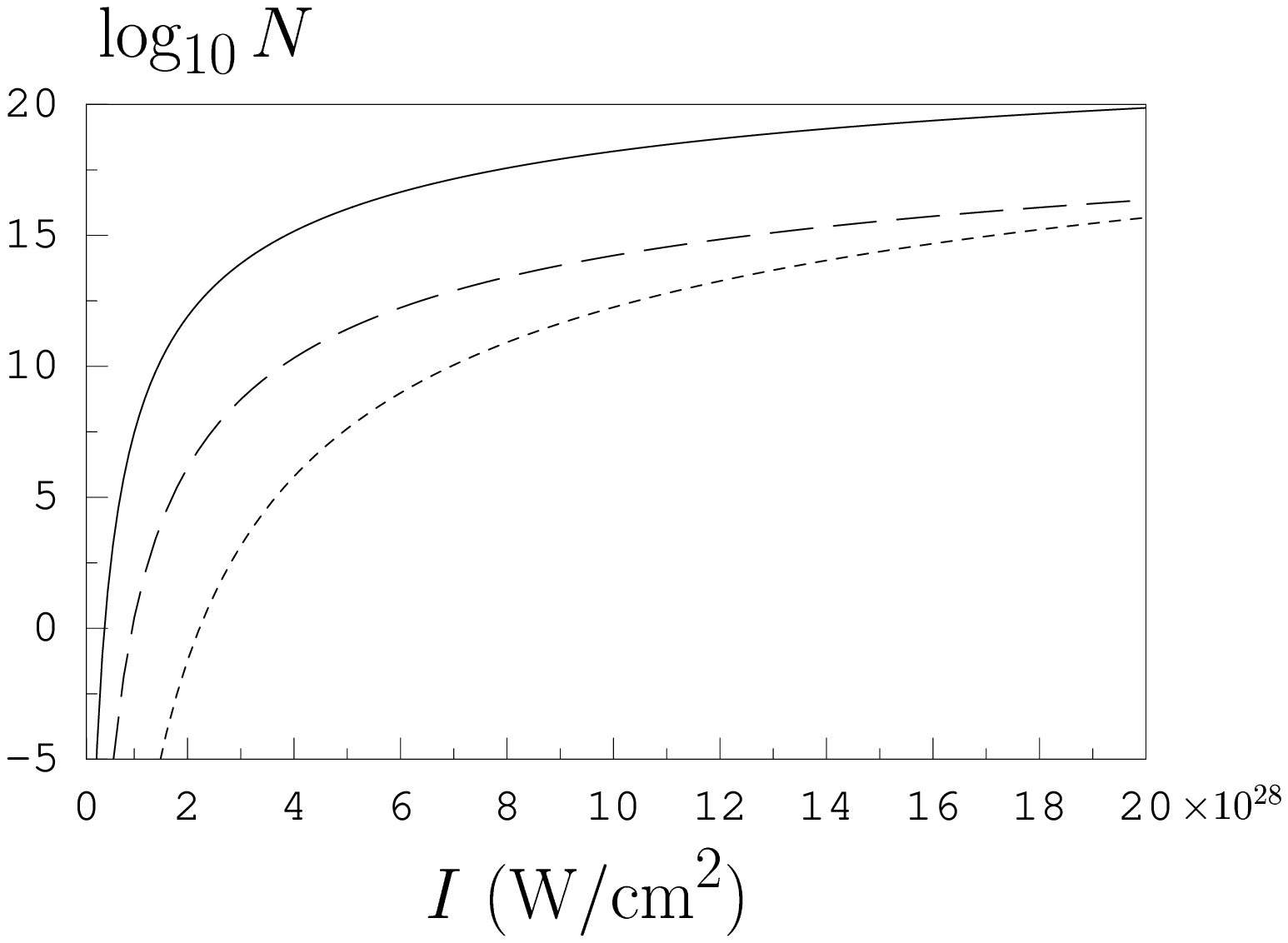} \caption{ Dependence of the
number of created pairs versus laser pulse intensity for
$e$-polarized (solid curve), $h$-polarized (dashed curve), and for
equal mixture of $e$- and $h$-polarized pulses (dotted curve).
$\lambda=1 \mu m$, $\tau=10^{-14}$s, $\Delta=0.1$.}
\end{figure}

In Fig.~5 we show the dependence of number of created pairs on the
laser pulse intensity for $e$- and $h$-polarized waves. Since in
the case of the $h$-polarized wave pairs are created at the
periphery of the pulse, where the field amplitude is exponentially
suppressed, the number of pairs is several orders of magnitude
less than in the case of the $e$-polarized wave. The dotted line
in Fig~5 represents the number of pairs created by an equal
mixture of $e$- and $h$-polarized pulses. We see that the number
of created pairs by the mixture is several orders of magnitude
less than even in the case of the $h$-polarized pulse. This
happens due to nonlinear dependence of invariants $\mathcal{E}$
and $\mathcal{H}$ on the field strengthes $\mathbf{E}$ and
$\mathbf{H}$. We have calculated the number of pairs created by an
arbitrary mixture of $e$- and $h$-polarized pulses and have found
that $e$ type of polarization is the optimal configuration of the
focused laser pulse for observation of the pair creation effect.

The numbers of pairs $N_{e,h}$ created by $e$- and $h$-polarized
waves for different values of $I$ and $\Delta$, as well as the
estimated values of $N_e$ for $e$-polarized wave, are given also
in the TABLE \ref{<Table>}. We have limited the values of
intensity in the TABLE \ref{<Table>} by $I=5\cdot 10^{28}$W/cm$^2$
since at larger intensities the exploited method of calculations
ceases to be valid.

\begin{table}[tbp]
\begin{tabular}[b]{||c||c||c|c||c||c||}
\hline $I$, W/cm$^2$ & $E_0/E_S$ & $N_e$, $\Delta=0.1$ &  estimate
& $N_e$, $\Delta=0.05$ & $N_h$, $\Delta=0.1$
\\ \hline $2\times 10^{27}$ & $0.11$ & $4.0\times
10^{-11}$ &  $3.4\times 10^{-12}$ & $4.6\times 10^{-42}$ &
$9.6\times 10^{-23}$
\\ \hline $4\times 10^{27}$ & $0.16$ & $0.09$
& $0.07$ & $6.8\times 10^{-24}$ & $1.5\times 10^{-10}$\\ \hline
$5\times 10^{27}$ &
$0.18$ & $24$ & $30$ & $3.11\times 10^{-19} $ & $2.45\times 10^{-8} $\\
\hline $6\times 10^{27}$ & $0.20$ & $1.5\times 10^3$ & $2.7\times
10^3$ & $8.8\times 10^{-16}$ & $4.4\times 10^{-5}$
\\ \hline
$8\times 10^{27}$ & $0.23$ & $5.3\times 10^5$ & $1.6\times 10^6$ &
$6.4\times 10^{-11}$ & $8.6\times 10^{-2}$
\\ \hline $1\times 10^{28}$ & $0.25$ & $3.0\times
10^7$ &  $1.2\times 10^8$ & $1.4\times 10^{-7}$ & $16$ \\ \hline
$2\times 10^{28}$ & $0.36$ & $8.0\times 10^{11}$ &  $7.1\times
10^{12}$ & $32$ & $8.5\times 10^6$
\\ \hline
$5\times 10^{28}$ &
$0.56$ & $1.0\times 10^{16}$ & $1.6\times 10^{17}$ & $1.3\times 10^9 $
 & $1.6\times 10^{12} $\\
\hline
\end{tabular}
\caption{\label{<Table>}The number $N_{e,h}$ of produced pairs for
different values of laser pulse intensity and parameter $\Delta$
for $e$- and $h$-polarized waves.}
\end{table}

One can see from the TABLE \ref{<Table>} that the estimates
obtained in the previous section agree rather well with the
results of numerical calculations. The estimates appeared to be
especially successful at $I\sim \tilde{I}\sim 5\cdot
10^{27}$W/cm$^2$. However we underestimate (overestimate) the
number of created pairs at lower (higher) intensities the estimate
yields approximately one order of magnitude. This is explained by
the fact that the higher is intensity, the sharper is the peak of
the integrand in Eq.~(\ref{NT}). This means that the higher is
intensity, the smaller is the effective radius $R_{eff}$ of the
area providing the principal contribution into the integral
(\ref{NT}). At $I\sim\tilde{I}\quad R_{eff}\sim R$, while at
$I<\tilde{I}\quad R_{eff}> R$ and at $I>\tilde{I}\quad R_{eff}<
R$. This point has not been taken into account by our method of
estimation.

\section{Conclusions}

In the present paper we have studied the effect of pair creation
by a circularly polarized focused laser pulse in vacuum. To
describe the focused laser pulse we have used a 3-dimensional
model of electromagnetic field based on an exact solution of
Maxwell equations \cite{NB}. It was shown that the number of
created pairs strongly depends on configuration of the
electromagnetic field in laser focus and on dimensionless focusing
parameter $\Delta$.

Pairs are created most effectively by a $e$-polarized pulse.
Indeed, for the $e$-polarized laser pulse with $\lambda=1\mu m,
\tau=10 fms$, and $\Delta=0.1$, the effect becomes observable at
intensity $I\approx 5\cdot 10^{27}$W/cm$^2$, while for the
$h$-polarized pulse with the same set of parameters only at
$I\approx 10^{28}$W/cm$^2$. In both cases the pair creation
process begins at intensities essentially less than the
characteristic value $I_S=5\cdot 10^{29}$W/cm$^2$. It is worth
noting that the peak value of the electric field in the focus in
both cases is less than the critical QED value $E_S$. As it was
mentioned in the preceding section, this is explained by a very
large value of the effective laser pulse 4-volume, where pairs are
effectively created, in comparison with the characteristic Compton
4-volume, $l_c^4/c$. One can see from the TABLE \ref{<Table>} that
for the $e$-polarized pulse with the same $\lambda$ and $\tau$ but
with $\Delta=0.05$ the effect becomes observable only at intensity
$I\approx 2\cdot 10^{28}$W/cm$^2$.

A very important consequence of our investigation is existing of a
natural physical limit for attainable focused laser pulse
intensities. This limit is posed by the effect of pair creation
and for the circularly $e$-polarized laser pulse with
$\lambda=1\mu m, \tau=10 fms$, and $\Delta=0.1$ is approximately
$0.3I_S$, see Sec.~III.

\acknowledgments

Authors are grateful to participants of Scientific Session
MEPhI-2004, Moscow, January 2004, for helpful discussion of the
results. This work was supported in part by the Russian Fund for
Fundamental Research under projects 03-02-17348 and 01-02-16850,
the Federal Program of the Russian Ministry of Industry, Science
and Technology grant No 40.052.1.1.1112 and by the Russian
Ministry of Education project No 1618.


\begin{thebibliography}{99}
\bibitem{Saut}  F. Sauter, Z. Phys. {\bf 69}, 742
(1931); {\bf 73}, 547 (1931).
\bibitem{el-pos}  W. Heisenberg and H. Euler, Z. Phys. {\bf 98}, 714
(1936).
\bibitem{Schwinger}  J. Schwinger, Phys. Rev. {\bf 82}, 664
(1951).
\bibitem{B-I}  E. Brezin and C. Itzykson, Phys. Rev. D {\bf 2}, 1191
(1970).
\bibitem{3}  V. S. Popov, JETP Lett. {\bf 13}, 185 (1971); Sov. Phys.
JETP {\bf 34}, 709 (1972).
\bibitem{4}  V. S. Popov, JETP Lett. {\bf 18}, 255 (1973); Sov. J. Nucl.
Phys. {\bf 19}, 584 (1974).
\bibitem{NN}  N. B. Narozhny and A. I. Nikishov, Sov. Phys. JETP {\bf 38}
, 427 ( 1974).
\bibitem{7}  V. M. Mostepanenko and V. M. Frolov, Sov. J. Nucl. Phys.
{\bf 19}, 451 (1974).
\bibitem{8}  M. S. Marinov and V. S. Popov, Fortschr. Phys. {\bf 25}, 373
(1977).
\bibitem{9}  A. A. Grib, S. G. Mamaev, and V. M. Mostepanenko, {\it %
Vacuum Quantum effects in strong fields} (Energoatomizdat, Moscow,
1988).
\bibitem{Ringwald}  A. Ringwald, Phys. Lett. B {\bf 510}, 107
(2001); arXiv: hep-ph/01112254, hep-ph/0304139.
\bibitem{Popov}  V. S. Popov, JETP Lett.{\bf 74}, 133 (2001); Phys. Lett.,
{\bf A298}, 83 (2002); JETP {\bf 94}, 1057 (2002).
\bibitem{Bula}  C. Bula, {\it et al}., Phys. Rev. Lett. {\bf 76}, 3116 (1996).
\bibitem{Burke} D.L. Burke, {\it  et al.}, Phys. Rev. Lett. {\bf 79}, 1626 (1997).
\bibitem{25}  F. V. Bunkin and I. I. Tugov, Dokl. Akad. Nauk SSSR {\bf 187%
}, 541 (1969).
\bibitem{26}  G. J. Troup and H. S. Perlman, Phys. Rev. D {\bf 6}, 2299
(1972).
\bibitem{Mourou}  T. Tajima, G. Mourou, Phys. Rev. ST-AB {\bf 5}, 031301 (2002);
arXiv: physics/0111091.
\bibitem{NB} N. B. Narozhny, M. S. Fofanov, JETP {\bf 117}, 867
(2000).
\bibitem{NF} N. B. Narozhny, M. S. Fofanov, Phys. Lett {\bf A 295}, 87
(2002).
\bibitem{MLM} G. Malka, E. Lefebvre, and J.L. Miquel, Phys. Rev. Lett.
{\bf 78}, 3314 (1997).
\bibitem{BW} M. Born and E. Wolf, {\it{Principles of Optics}} (Pergamon Press,
New York, 1964)

\end{thebibliography}
\end{document}